\documentclass{article}

    \PassOptionsToPackage{numbers, compress}{natbib}

\usepackage{graphicx}

    \usepackage[preprint]{neurips_2022_ml4ps}

\usepackage[utf8]{inputenc} 
\usepackage[T1]{fontenc}    
\usepackage{hyperref}       
\usepackage{url}            
\usepackage{booktabs}       
\usepackage{amsfonts}       
\usepackage{nicefrac}       
\usepackage{microtype}      
\usepackage{xcolor}         

\title{Emulating cosmological growth functions with B-Splines}

\author{%
    Ngai Pok Kwan\\
    Department of Physics\\
    The Chinese University of Hong Kong\\
    Center for Computational Astrophysics\\
    Flatiron Institute, NY \\
    \texttt{1155158081@link.cuhk.edu.hk} \\
     \And
    Chirag Modi  \\
    Center for Computational Astrophysics \\
    Center for Computational Mathematics \\
    Flatiron Institute, NY \\
    \texttt{cmodi@flatironinstitute.org} \\
     \And
    Yin Li  \\
    Center for Computational Astrophysics \\
    Center for Computational Mathematics \\
    Flatiron Institute, NY \\
    \texttt{yinli@flatironinstitute.org} \\
    \And
    Shirley Ho \\
    Center for Computational Astrophysics \\
    Flatiron Institute, NY \\
    \texttt{sho@flatironinstitute.org} \\
}

\begin{document}

\maketitle

\begin{abstract}
In the light of GPU accelerations, sequential operations such as solving ordinary differential equations can be bottlenecks for gradient evaluations and hinder potential speed gains. 
In this work, we focus on growth functions and their time derivatives in cosmological particle mesh simulations and show that these are the majority time cost when using gradient based inference algorithms. 
We propose to construct novel conditional B-spline emulators which directly learn an interpolating function for the growth factor as a function of time, conditioned on the cosmology. 
We demonstrate that these emulators are sufficiently accurate to not bias our results for cosmological inference and can lead to over an order of magnitude gains in time, especially for small to intermediate size simulations. 
\end{abstract}

\section{Introduction}

Field level inference for cosmological analysis
simulates the survey observations data at the level of full field by starting all the way from the initial density distribution at the beginning of the Universe and then evolving dark matter particles under gravity with Particle Mesh (PM) N-Body simulations. 
The goal then is to infer the cosmological parameters along with the initial conditions at all points in the Universe.
This challenging high dimensional inference relies on using differentiable simulations \citep{borg, flowpm, madlens, pmwd} and coupling them with gradient based algorithms such as Hamiltonian Monte Carlo (HMC) \citep{Neal11}.
Due to the iterative nature of these algorithms, it is crucial for these simulators to be simultaneously fast and accurate. 

Recent works have used advances in automatic differentiation libraries such Tensorflow and Jax to build these simulators like FlowPM \citep{flowpm} and pmwd \citep{pmwd}.
An added advantage of this is that the simulations can now exploit efficient GPU parallelizations and accelerations for significant speed-ups.
However this has inadvertently made other sequential operations a bottleneck to fully realize potential gains. One such example is solving ordinary differential equations (ODE) wherein gradient calculation has to backpropagate through all the sequential computations of the integrator \citep{neuralode}. 
In cosmological simulations, the growth factor of density fields and distance functions are estimated by solving a system of ODEs as a function of time. 
We find that for small to intermediate PM simulations, backpropogating through growth function ODE can be the majority time cost. 
Thus in this work, we seek to replace this ODE solution with trained emulators. 

In the past couple years, many works have built emulators for time intensive  operations in cosmological analysis pipeline \citep{cosmopower, DeRose22}. 
One way these works learn the emulator is by training a multi-layer perceptron (MLP) to fit the quantity of interest at some fixed points in the domain. Then during analysis, they construct an interpolating function through these points. However, it is redundant to query an MLP and then the interpolating function. For accurate interpolation, these output points also need to be densely sampled, making output of MLP high dimensional and challenging to learn.
An alternative approach is to learn the coefficients for principal components basis but that limits the accuracy based on the number of principal components used. Depending on the analysis, it also does not necessarily solve the interpolation redundancy.

In this work, we take a different approach and learn the emulator by directly learning the components of an interpolating function. Specifically, we learn a B-Spline emulator \citep{deboor} that can be parameterized by a small number of knot points and weights to model a smooth, high order differentiable function. 
We begin by setting up the growth function ODE in section \ref{sec:growth} and follow it with discussing our emulator in section \ref{sec:emulator}. In section \ref{sec:results}, we show that the emulator meets desired accuracy and quantify achieved gains before concluding with a brief discussion in section \ref{sec:discussion}.

\section{Growth function in cosmology}
\label{sec:growth}
Cosmological simulations evolve dark matter over-density field under gravitational force in an expanding universe. This evolution is governed by a solving a system of coupled non-linear partial differential equations called Vaslov Poisson equations. However at the linear order and after making a number of simplifying assumptions not detailed here for brevity, this evolution is governed by the following ODE \citep{Peebles80}
\begin{equation}
    \frac{\partial^2 \delta(x,t)}{\partial t^2} + 2{H(t)}\frac{\partial \delta(x, t)}{\partial t}  = \frac{3}{2}\Omega_m(t) H_0^2\delta(x, t)
\label{eq:evolve}
\end{equation}
where ${H}(a) = H_0(\frac{\Omega_m}{a^3} + \Omega_{\Lambda})^{0.5}$ is the Hubble parameter, $H_0$ is the Hubble constant, $\Omega_m(a) = \frac{\Omega_m}{a^3}\frac{H_0^2}{H^2(a)}$ is the matter density,  $\Omega_{\Lambda}$ is the dark energy density and $a$ is the scale factor of the Universe, defined as $H(a) = \frac{1}{a}\frac{d a}{dt}$.
For simplicity, in the following we use $a$ as a measure of time instead of $t$ since that is the convention in the particle mesh simulations of interest here.

At linear order, the evolution of the density field can be decoupled into spatial and time contributions by writing $\delta(x, a) = D_1(a)\delta(x, a_0)$ for some ``reference" time $a_0$ and $D_1(a)$ called the linear growth factor \citep{Hamilton00}. Then the time dependence of the growth factor is 
\begin{equation}
    a^2\frac{d^2 D_1(a)}{da^2} + \Bigg(\Omega_\Lambda(a) + \frac{\Omega_m(a)}{2} + 2 \Bigg)a\frac{d D_1(a)}{d a}  = \frac{3}{2}\Omega_m(a)D_1(a)
\label{eq:d1}
\end{equation}
Being a second order equation, this has two solutions. One of them increases with time and is referred to as growing mode ($D_1^+$) while the other decays with time and hence is generally ignored in the simulations. 
At higher orders in the Taylor expansion of the density field, we can define a similar growth function at the second order for $\delta^{(2)}(x,a)$ that follows the ODE
\begin{equation}
    a^2\frac{d^2 D_2(a)}{da^2} + \Bigg(\Omega_\Lambda(a) + \frac{\Omega_m(a)}{2} + 2 \Bigg)a\frac{d D_2(a)}{d a}  = \frac{3}{2}\Omega_m(a)\big[D_2(a) - (D_1^+(a))^2\big]
\label{eq:d2}
\end{equation}

\subsection{Growth function in PM simulation}
Particle mesh simulations evolve this density field by computationally evolving dark matter particles under gravity. 
At every time step, these simulations consist of three consecutive operations- \emph{kick} step updates the particle momentum, \emph{drift} step which updates the positions of the particles and \emph{force} step which estimates the force on each particle in this new configuration. In FastPM simulations \citep{Feng2016}, the growth factors determine the scaling of the particle displacement in the drift step. For a time step from $a_0$ to $a_1$, the displacement is given by 
\begin{equation}
    x(a_1) = x(a_0) + \frac{H_0}{a_r H(a_r)} \frac{D(a_1) - D(a_0)}{dD/da |_{a_r}}p(a_r)
\end{equation}
where $a_r$ is a reference time between $a_0$ and $a_1$ and $p$ is the particle momentum.

\begin{figure}
  \centering
 \includegraphics[width=1.\columnwidth]{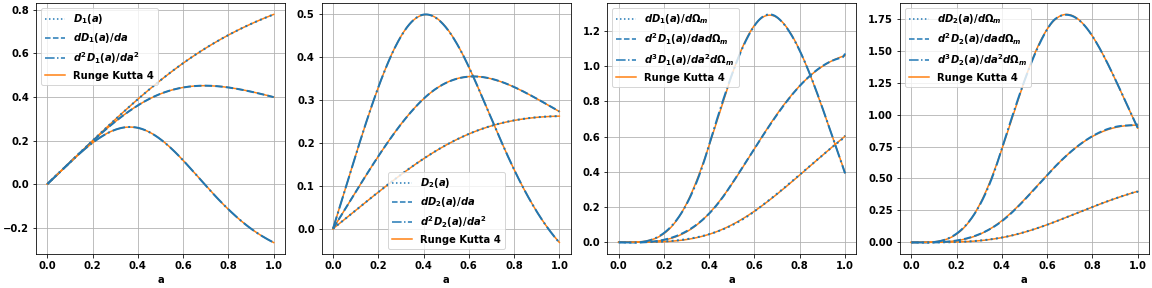}  
  \caption{At $\Omega_m=0.3$, the output of emulator/ML (blue lines) compared to the true value (orange lines). The first two columns show the first and second order growth function ($D_1$ and $D_2$). The third and fourth columns show the derivatives of $D_1$ and $D_2$ with respect to $\Omega_m$. The function values (dotted) are plotted together with the two derivatives with respect to $a$ (dashed and dash-dotted).}
  \label{fig:functions}
\end{figure}

Thus at every drift step, we need to estimate the growth factor at the beginning and the end time of the step, as well as its time derivative at a reference point in between.
While not shown here (but see \cite{Feng2016}), the momentum update similarly depends on two time derivatives, $dD/da$ and $d^2D/da^2$ of the growth function. We show all these functions for a fiducial cosmology in Figure \ref{fig:functions}.

\section{B-spline Emulator}
\label{sec:emulator}
We are interested in constructing an emulator for the growth factor and its derivatives as a function of cosmology. In the standard $\Lambda$CDM model, Eq. \ref{eq:d1} tells us that the growth factor is only a function of $\Omega_m$ since $\Omega_m + \Omega_{\Lambda}=1$. Hence our emulator has only one input.


We learn a B-spline (or basis spline) function \citep{deboor} to model the growth function. 
B-splines are powerful since they are the maximally differentiable interpolative basis function and any spline function of a given degree $n$ can be expressed as a unique linear combination of B-splines of that degree. 
B-splines are defined by the number of interior `knots' N. Thus let $t_0,t_1,...,t_N,t_{N+1}$ be a non-decreasing sequence of knots. These knots are augmented by repeating exterior knots $t_0$ and $t_{N+1}$ $n$ times and for each augmented knot $t_i$, a set of basis functions $B_{i,j},\, \forall j=0,1,...,n$ is defined recursively from $j=0$ to $n$.
A B-spline function is then the linear combinations of these basis functions with weights $\beta_i$
\begin{equation}
    B(x) = \sum_{i=0}^{i=N+n} \beta_i B_{i,n}(x), \quad x \in [t_0, t_{N+1}]
\end{equation}

Thus a B-spline function $B(x)$ is completely defined given a sequence of knots $t_i$ and weights $\beta_i$.
In our emulator, we train an MLP to predict these knots and weights for an input cosmology.

\section{Results}
\label{sec:results}
We train our emulator as a function $\Omega_m$ to predict the knot points and weights of cubic $(n=3)$ B-spline function. The emulator works in the range of $\Omega_m\in [0.1, 0.5]$. $a$ is in the range of 0 to 1. 1000 random values of $\Omega_m$ from 0.05 to 0.55 are generated by uniform sampling for training. There are 256 points evenly distributed in the range of $a$, including 0 and 1. The training, validation and testing data sets are of size 800, 100 and 100 respectively.

\begin{figure}
  \centering
 \includegraphics[width=1.\columnwidth]{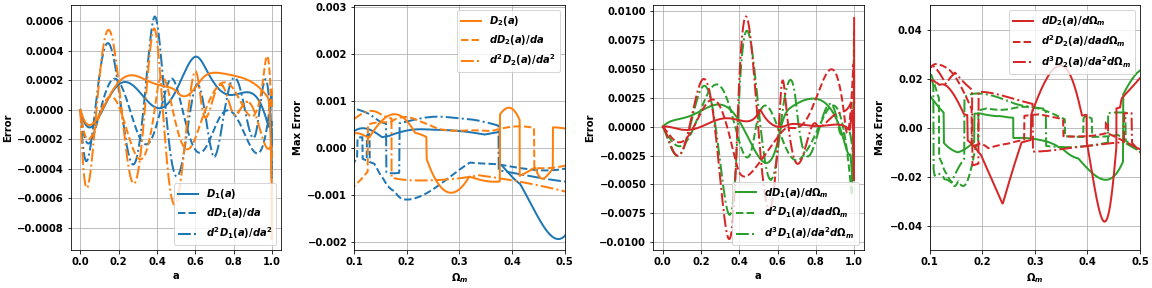}  
  \caption{(Subplot 1 and 3) Error of growth functions at $\Omega_m=0.3$ and $a\in[0, 1]$.\\
  (Subplot 2 and 4) Maximum error of gradients across $a\in[0, 1]$ for $\Omega_m\in[0.1, 0.5]$. }
  \label{fig:emulate}
\end{figure}

We find that using only 8 knots for $a\in[0, 1]$ is sufficient for achieving sub-percent accuracy. For the MLP architecture, there is first an input layer. The output of the input layer is duplicated and fed into two parts in parallel, each having a hidden layer and an output layer. The two parts account for the calculation of knot points and weights respectively. The input layer and two hidden layers have 64 neurons each. The neurons of output layers depend on the number of knots. In our case, the MLP outputs 8 knot points and the corresponding weights. One knot is fixed at $a=0$.
We implement our emulator as part of pmwd code \citep{pmwd}.
We begin by showing the accuracy of our emulator in Figure \ref{fig:emulate}.

We show errors in value and gradient for a reference cosmology of $\Omega_m=0.3$. In addition, for every cosmology, we also show the maximum error among all $a\in[0, 1]$  in the last two panels. The error in value is always less than 0.002 and in gradient is always less than 0.05. Next, we show that this level of agreement does not impact the accuracy of our inference.

\begin{figure}
  \centering
 \includegraphics[width=1.\columnwidth]{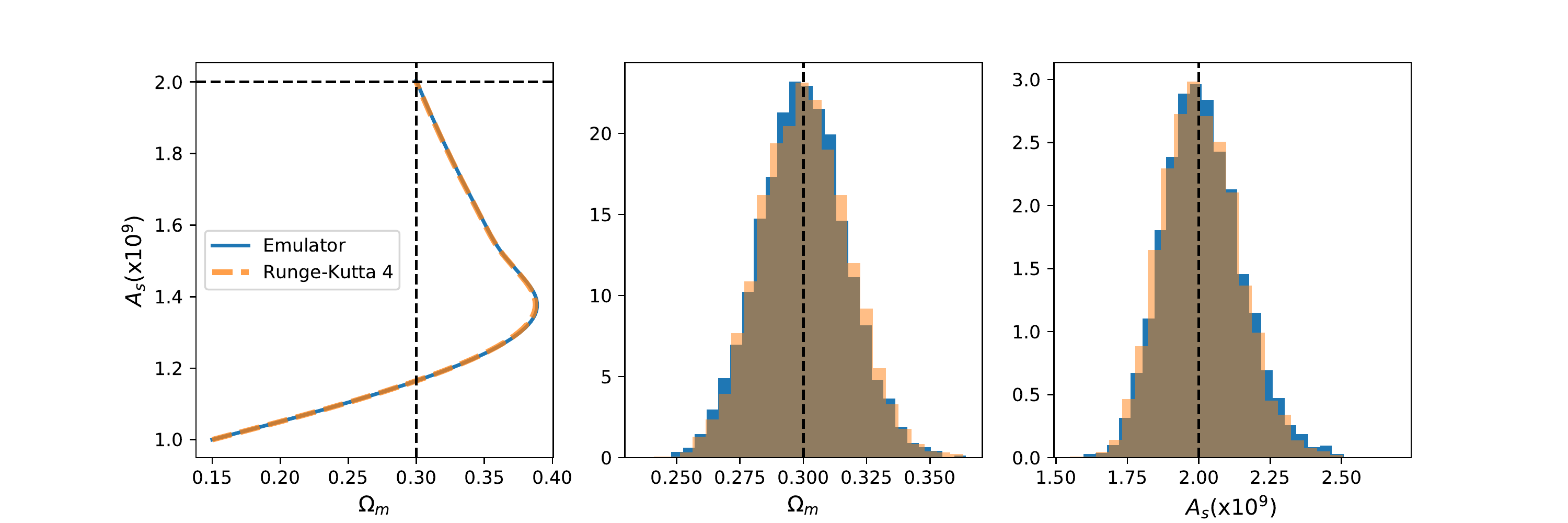}  
  \caption{(Left) Trajectory of MAP optimization (middle, right) Marginal posteriors sampled with HMC for a toy field level inference problem. Dashed black lines are true parameters. }
  \label{fig:opt}
\end{figure}

As noted in the beginning, the objective of using differentiable PM simulations is to be able to access the gradients of cosmology parameters for field level inference. We mock this pipeline with a simulated data to infer $\Omega_m$ and $A_s$ (scalar amplitude) parameters.
Figure \ref{fig:opt} shows the results for two gradient based approaches when using our emulator versus ODE formulation for growth function. 
In the left panel, we show the trajectory followed when doing optimization to find the MAP (maximum-a-posteriori) estimate of the cosmology parameters.
The last two panels show the marginal posteriors sampled with HMC for the same problem. 
The agreement between our emulator and ODE solution validates that the accuracy of the emulator and its gradients is sufficient for correct inference.

\begin{figure}
  \centering
 \includegraphics[width=1.\columnwidth]{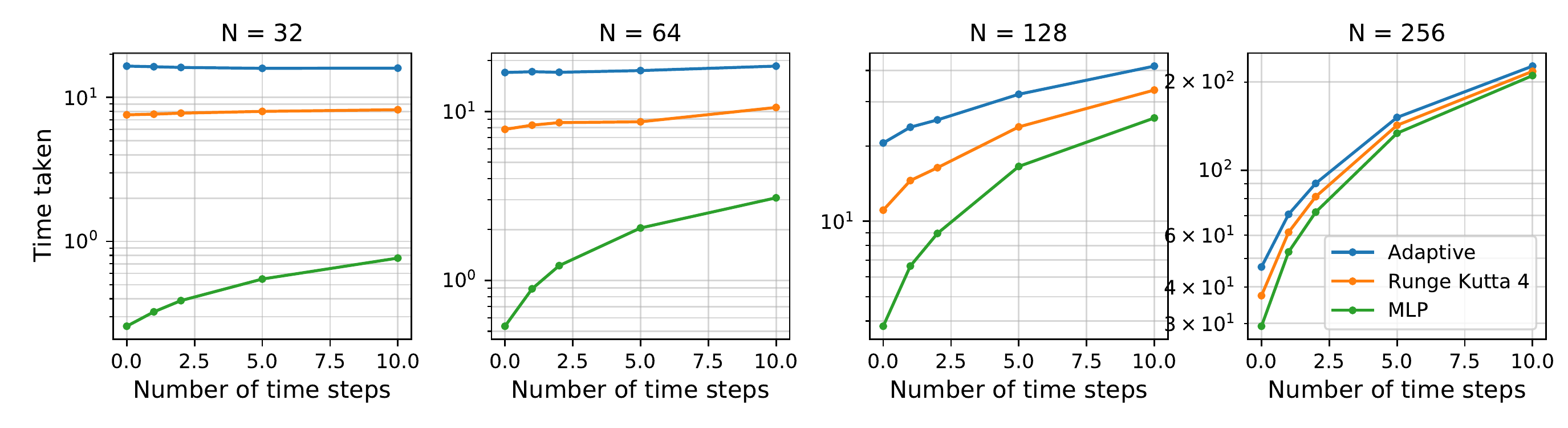}  
  \caption{Time for 100 gradient evaluations with respect to cosmology parameter for simulations with of different number of particles (N). 0 time step corresponds to setting initial conditions.}
  \label{fig:timings}
\end{figure}

Finally, in Figure \ref{fig:timings}, we show the timings for gradient evaluation when using our emulator versus the adaptive Dormand-Prince ODE solver and Runge Kutta (order 4, solved at 256 points in $a\in[0, 1]$).
For $N < 128$, increasing the simulation size does not change the time cost, demonstrating that ODE poses the bottleneck. In this case, we can gain up to an order of magnitude in time with the emulator. For larger simulations, PM cost starts to dominate as expected and we gain up to a factor of 2.

\section{Discussion}
\label{sec:discussion}
In this work, we demonstrate that for highly optimized PM simulations exploiting GPU accelerations, sequential operations such as solving ODE for growth function can be the majority time cost for gradient evaluations.
We replace them with novel B-Spline emulators which can lead to an order of magnitude gains for small to intermediate simulations. 
While the gains are less remarkable for large $N\geq256$ simulations, it is important to note that most of the methodology development is done on the small simulations with large runs primarily done only at the final analysis. 
Hence these gains will still lead to non-trivial time savings in developing novel methods for cosmological inference.
We also anticipate similar bottleneck for distance calculation in weak lensing simulation and plan to explore that in the future. In the next work, we plan to extend the emulator to account for other cosmological parameters, such as the curvature $\Omega_k$, and dark energy parameters $w_0$ and $w_a$.

\typeout{}
\bibliographystyle{JHEP}
\bibliography{biblio}




\end{document}